\documentclass{article}

\usepackage[final]{neurips_2019}

\usepackage[utf8]{inputenc} 
\usepackage[T1]{fontenc}    
\usepackage{hyperref}       
\usepackage{url}            
\usepackage{booktabs}       
\usepackage{amsfonts}       
\usepackage{nicefrac}       
\usepackage{microtype}      
\usepackage{eurosym}
\usepackage{amsmath}
\usepackage{graphicx}

\title{FDL: Mission Support Challenge}

\author{%
  Lu\'is F. Sim\~oes, Ben Day, Vinutha M. Shreenath, Callum Wilson\\
  Chris Bridges, Sylvester Kaczmarek \& Yarin Gal\\
  Mission Support Team, \\
  Frontier Development Lab. \\
  \texttt{ben.day@cl.cam.ac.uk, callum.j.wilson@strath.ac.uk} \\
}

\begin{document}

\maketitle

\section{Introduction}
\paragraph{The program} The Frontier Development Lab (FDL) is a National Aeronautics and Space Administration (NASA) machine learning program with the stated aim of conducting \textit{artificial intelligence research for space exploration and all humankind} with support in the European program from the European Space Agency (ESA). Interdisciplinary teams of researchers and data-scientists are brought together to tackle a range of challenging, real-world problems in the space-domain. The program primarily consists of a sprint phase during which teams tackle separate problems in the spirit of \textit{coopetition}\footnote{\textit{Coop}eration + comp\textit{etition} = \textit{coopetition}. Cooperation in providing assistance where we have specialist knowledge, competition in spurring each other on by example and the desire to `be the best'.}. Teams are given a problem brief by real stakeholders and mentored by a range of experts. With access to exceptional computational resources, we were challenged to make a serious contribution within just eight weeks.

\paragraph{Our challenge} Stated simply, our team was tasked with producing a system capable of scheduling downloads from satellites autonomously. Scheduling is a difficult problem in general, of course, complicated further in this scenario by ill-defined objectives \& measures of success, the difficulty of communicating tacit knowledge and the standard challenges of real-world data. Taking a broader perspective, spacecraft scheduling is a problem that currently lacks an intelligent solution and, with the advent of mega-constellations, presents a serious operational bottleneck for the missions of tomorrow.

\section{Challenge description}
\textit{Cluster-II} is a constellation of four spacecraft monitoring the Earth's magnetosphere and its interaction with the solar-wind. In the $20^{th}$ year of a mission originally planned to last just 2 years, the data collected has contributed to over 3500 published papers and has been an example of international cooperation in space with a NASA payload on-board and a joint mission coordinated with the Chinense National Space Administration (CNSA) constellation \textit{Double Star}.

Each of the four spacecraft in the constellation (code-named Salsa, Samba, Rumba \& Tango) continuously monitor their surroundings with a variety of scientific equipment, rewriting previous observations in a looped storage system -- the oldest recording is overwritten by the latest. It is the primary objective of scheduling to ensure that no data is overwritten before it has been retrieved from the spacecraft. Downlinking from a spacecraft to a ground-station may only occur when the spacecraft is visible (in-the-sky of the station) with the bitrate of the transfer determined by factors including the orbital distance, relative orientation, antenna quality, and local environmental \& weather effects. A booking is referred to as a \textit{pass} and the time in which a booking is possible is the \textit{visibility}. There are also \euro-cost considerations: how long to book use of the ground-station and how to best coordinate with shift schedules to minimize overtime. And, as the craft fly in close proximity to one another, possible periods of downlink will overlap and trade-offs will need to be made between them. However, when the angular separation of the spacecraft is sufficiently low, a single ground-station is able to receive data from two craft at once. This rich tapestry of objectives, constraints and considerations (not all of which are able to be detailed here), together with nearly two decades of human-made examples, makes for a challenging, yet tantalisingly solvable, problem.

\paragraph{Prior work} The operations team running \textit{Cluster-II}, in collaboration with the AI teams at ESA, have made efforts in the past to automate the scheduling process. Most notably a constraint-solver based approach, TIAGO, \cite{Faerber2016Cluster-II:Scheduling}, is able to produce schedules that are physically plausible and meet basic mission objectives, but the operators considered the solutions produced to be unsuitable due to concerns about the brittleness of the resulting schedules (coming dangerously close to losing data, having little recourse if things were to not go as planned). The system also creates solutions that are strikingly different from those a human would produce and, presumably, this unfamiliarity is also unfavourable.

\section{Approach}
We spent a week at ESA's operations centre (ESOC) alongside the team running the mission. During this time we were able to observe operators creating schedules, allowing us to gain an understanding of the strategies employed by experts and the subtleties of the problem. It also became clear as to why the constraint solver had been considered insufficient: much of what the most skilled operators do is based on tacit knowledge, their expertise, that can not easily be described as a series of constraints. Though much of what we observed could not be formalised, many key objectives can be stated simply. For example, the \euro-cost of booking passes at ground-stations for different lengths of time is well established: for less than an hour, you pay for one hour; for more than an hour, you pay for the total time plus one additional hour. We were able to formalise nine scores capturing aspects of cost-efficiency, signal quality, storage fill level, shift-alignment, and other niche considerations \textbf{detailed in appendix \ref{appendix:scores}}.

The most significant challenges were in preparation, becoming `\textit{ML-ready}'. Drudging through disparate sources trying to unpick connections, we evaluated precedence to establish the most accurate picture presented by the many data sources. There were many difficulties and more hours were spent here than in any other single task. In addition to data preparation, we also developed tools to simulate the scheduling context and a scheduling environment. The possible connection quality during a visibility is called the \textit{link-budget}. We wrote a calculator using the \texttt{linkpredict} library, \cite{linkpred}, able to use web-scraped TLE format descriptions of the orbits or absolute positions, where provided, that allows the link-budget to be calculated at any time, past or future. We also implemented a fill-level simulator, able to extrapolate fill-level curves and update the curves based on booked passes -- the environment for a scheduler. \textbf{Examples of the output of these modules can be found in appendix \ref{appendix:simulators}.}

In order to surpass the constraint solver, we attempted to model the decision-making of human operators for use as an additional evaluation metric. The argument being that if we can accurately judge the human-likeness of a schedule, or a decision, we may be able to train models that conform to the less easily elicited preferences of operators. Using the historic schedule data, we trained simple decision tree models to predict whether an operator would book a pass in a given visibility. To our great surprise, an ensemble of 50 trees of max-depth 3, given only a highly simplified context, was able to achieve an AUROC of 0.79 in a 10-fold cross-validation evaluation.

\paragraph{Scheduler} With the data prepared and scores \& simulator implemented, we were able to produce a scheduler. Our method uses a beam-search, a kind of tree search, guided in the first iteration simply by scores but with the ability to incorporate the human model as an additional heuristic. \textbf{Details are provided in appendix \ref{appendix:models}.}

\section{Conclusion \& future work}
The promise shown by our approach was deemed sufficient by the \textit{Cluster-II} operations manager to be worth developing. The next stage would be making a full evaluation of the schedules produced by the system by expert operators to determine whether the output meets their standards and, if not, what can be learned from this attempt and transferred to the next solution.

\section*{Acknowledgements}
We would like to thank our colleagues and friends at ESA, Bruno Sousa, Simone Fratini, Alessandro Donati and the members of the \textit{Cluster-II} operations team for their patience, encouragement and support. We send our thanks the the designers from FDL, Renee Verhoeven \& Leo Silverberg, and the FDL leadership team, Jodie Hughes, Sarah McGeehan \& James Parr. We would also like to thank the sponsors of the FDL program, in particular the Google Cloud Platform. And finally to acknowledge the developers of the Desmos graphing calculator \citep{desmos} without which we would have needed a lot more paper. 

\bibliographystyle{plainnat}
\bibliography{references,extra}

\clearpage

\appendix
\section{Scores}
\label{appendix:scores}
The scoring functions were each designed to capture some small, identifiable aspect of the operator objectives that, in combination, describe the explicit overall goals of the mission. We adopt names for the scores from the operators where provided. Each function maps to the closed interval $[0,1]$ with a worst-possible scenario assigned a score of 0, an ideal scenario given a score of 1 and better schedules receiving higher scores. The scores were also designed such that small changes in the schedule result in small changes in the score.

\begin{figure}
    \centering
    \includegraphics[width=\linewidth]{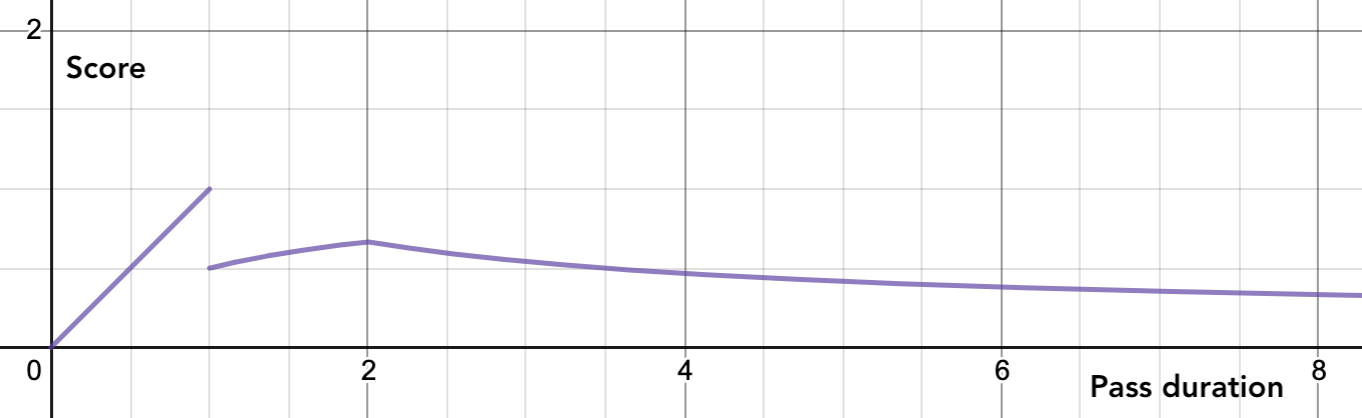}
    \caption{Pass cost-efficiency score example. In this example there are 10 units of data in storage (not shown), data is produced at 1 unit per unit time, and we download at 6 units per unit time. For passes less than 1 hour in length the efficiency increase linearly as the data downloaded is linear in time and the cost is fixed (so the numerator is increasing linearly and the denominator is constant). At 1 hour there is a step decrease in score corresponding to the jump in cost from $c(t\leq1)=t$ to $c(t>1)=t+1$ -- the numerator is constant but the denominator doubles. From the end of the first hour to the second the score again increases, proportional to $t/(t+1)$, asymptotically approaching 1. However, after 2 hours the storage is depleted ($10+2-(2\times6)=0$) and the download rate is throttled by the rate of production (1 unit per unit time). The efficiency now decreases as the download grows more slowly. This kink in the curve explains why we would want to \textit{either} book passes of 1 hour length \textit{or} fully deplete the storage. This is a strategy employed by the operators.}
    \label{fig:pass_cost}
\end{figure}

\paragraph{Pass cost-efficiency} Passes are scheduled to download data and are charged depending on how long the pass lasts. There are three possible download speeds used (high bit rate, low bit rate, zero) depending on the possible strength of connection (signal-to-noise). The operators are charged by the ground-stations based on the length of the pass non-linearly. In hours,
\begin{align*}
    \text{cost}(\Delta t) =
        \begin{cases}
            1 & \Delta t\leq1 \\
            \Delta t+1 &\text{otherwise}.
        \end{cases}
\end{align*}
For this score (and many that follow) we can simply define the maximal and minimal efficiencies and \textit{normalise} to this range as
\begin{align*}
    s' = \frac{s - \text{minimal}}{\text{maximal} - \text{minimal}}.
\end{align*}
The minimal efficiency is 0, downloading nothing during the pass, and downloading at high bit rate for 1 hour is maximally efficient. Thus the score
\begin{align*}
    S_{\text{PCE}} =  \frac{\text{download}}{\text{cost}(\Delta t)} \times \frac{\text{cost}(1 \text{hour})}{\text{fast download rate} \times \Delta t}.
\end{align*}
Figure \ref{fig:pass_cost} shows how the score varies in a toy example.

\begin{figure}
    \centering
    \includegraphics[width=\linewidth]{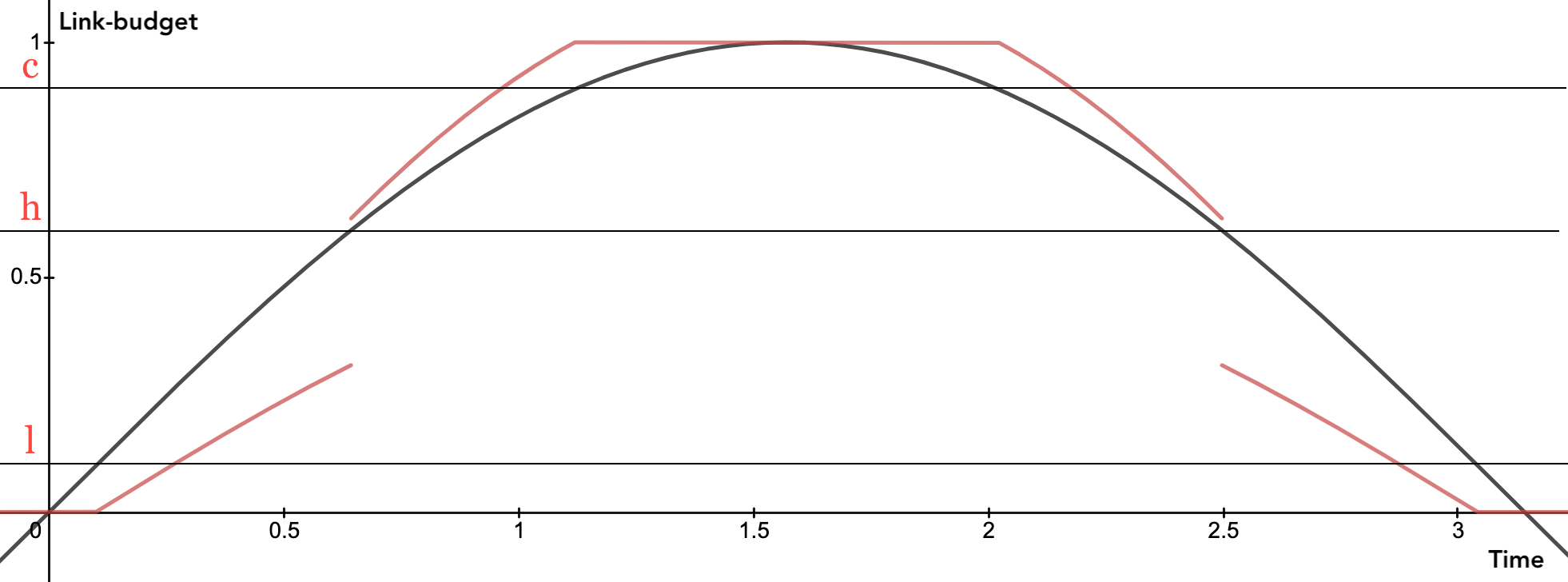}
    \caption{Quality over time. The black curve (a stand-in sinusoid) shows the link-budget, $q$. The red curve is the alignment quality $q'$. The horizontals mark the thresholds: where the black curve is below $l$ the red curve is 0; between $l$ and $h$ the curve is lower; between $h$ and $c$, higher; and above $c$ it maxes out at 1. In this example, if the pass started at $0.5$ and ended at $2.5$ the score would be $0.82$, but if it was between $1$ and $2$ the score would be over $0.99$ as the link budget is greater than the ceiling value for almost all of that interval.}
    \label{fig:align}
\end{figure}

\paragraph{Link-budget alignment} When booking a pass it is important not only to consider whether it is possible to download in a given rate, being over a signal-to-noise threshold, but also how far above the threshold. There are atmospheric effects that may cause the noise to increase resulting in a lessened quality of connection. In practice this means not booking passes during periods where the predicted signal-to-noise will hover around the threshold for a while. To score this we can consider how well a pass \textit{aligns with the link-budget}. The link-budget is already a real-valued function varying from all noise ($-\inf$) to all signal ($\inf$) but we only care about values between the low bit-rate threshold, $l$, and the effective ceiling, $c$, to improved signal strength (beyond which improvements aren't relevant). It is also preferable to be above the high bit-rate threshold, $h$. We first preprocess the link budget as
\begin{align*}
    q'= \begin{cases}
            0 & q\leq l \\
            \frac{q-l}{k(c-l)} & l \leq q < h \\
            \frac{q-l}{(c-l)} & h \leq q < c \\
            1 & c \leq q.
        \end{cases}
\end{align*}
where $k$ parameterises the improvement in quality from low to high (in our case $k=2$) and then to score a pass we integrate this quality of connection over the period
\begin{align*}
    S_{\text{LBA}}(q'(t),t_{\text{start}},t_{\text{end}})=\frac{\int_{t_{\text{start}}}^{t_{\text{end}}}q'(t)dt}{\int_{t_{\text{start}}}^{t_{\text{end}}}1dt} = \frac{\int_{t_{\text{start}}}^{t_{\text{end}}}q'(t)dt}{t_{end} - t_{start}}=\text{mean}(q'(t)).
\end{align*}
Figure \ref{fig:align} gives a visual example of how $q'$ varies and the scores produced.

\begin{figure}
    \centering
    \includegraphics[width=\linewidth]{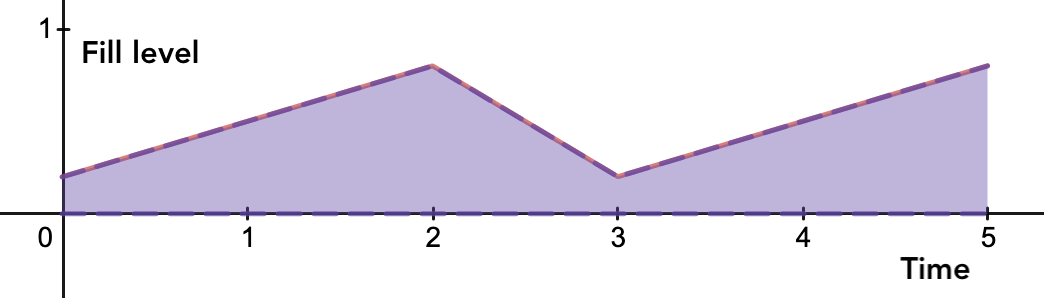}
    \caption{Fill level over time. The red line shows the fill level at a given time, $f(t)$, and the purple area is equal to the total fill, $F$. In this case the area under the curve is 2.5 with a time length of 5, giving a score of 0.5. The fill is, on average, half full.}
    \label{fig:fill}
\end{figure}

\paragraph{Fill level} The fill level should be kept low. The fill level at a given time, $f(t)$, is a real number between 0 (empty) and 1 (full). We measure the fill level over a period of time as the integral
\begin{align*}
    \text{total fill} = F = \int_{t_{start}}^{t_{end}}f(t)dt
\end{align*}
where the max fill, $F_m$, would be $t_{end} - t_{start}$. This renders the score
\begin{align*}
    S_{\text{FL}}\big(f(t),t_{start},t_{end}\big) = 1 - \frac{F}{F_m} = 1 - \text{mean}(f(t))
\end{align*}
depicted in figure \ref{fig:fill}. An additional precaution taken by the operators is to aim to keep the storage not just low but strictly lower than 75\% full. This makes the schedule more robust to missing passes (due to unpredictable technical issues either on the spacecraft or at the ground-station). This can be included by re-scaling the fill-level, most simply with a cut-off
\begin{align*}
    f'= \begin{cases}
            \frac{1}{\alpha}f & f<\alpha \\
            1 & \text{otherwise}.
        \end{cases}
\end{align*}
If we set $\alpha = 0.75$ then being 75\% full is considered to be as bad as being full. Smoothed alternatives and functions that have a similar tailing off for sufficiently low fill levels (`equally-good') can be easily imagined.

\paragraph{Fragmentation} The operators are concerned with how spread out passes are over a week. A schedule that is able to have passes occur in bursts is preferable to one that has passes spread evenly, primarily for the sake of scheduling shifts. This is referred to internally as \textit{fragmentation}. We considered two scores of fragmentation, an entropy based measure of clumpiness and the Fano factor which measures dispersion. In practice these scores strongly correlate, with subtle differences in the harshness of scoring at different parts of the range. Both scores work with the intervals between events. In the case of satellite operations, the start and end of passes are equivalent work and so are considered separate events. Where $x_0$ is the gap between the start of the considered time interval and the first event and $x_i$ is the gap between the $i^{th}$ and $(i+1)^{th}$ events, the entropic measure is
\begin{align*}
    S_{\text{F1}}(\text{intervals}) = 1 + \frac{\sum_{i=1}^{n+1}x_i\log(x_i)}{\log(n+1)}
\end{align*}
which is independent of the base (it is effectively changing base to $n+1$). The Fano factor score is simply the Fano factor
\begin{align*}
    S_{\text{F2}}(\text{intervals}) &= \frac{\sigma^2}{\mu} = \frac{\frac{1}{n}\sum_{i=1}^{n+1}(x_i-\mu)^2}{\frac{1}{n+1}} \\
    &= \frac{n+1}{n}\sum_{i=1}^{n+1}\big(x_i-\frac{1}{n+1}\big)^2.
\end{align*}

\section{Simulators}
\label{appendix:simulators}
\subsection{Link-budget}
\begin{figure}
    \centering
    \includegraphics[width=\linewidth]{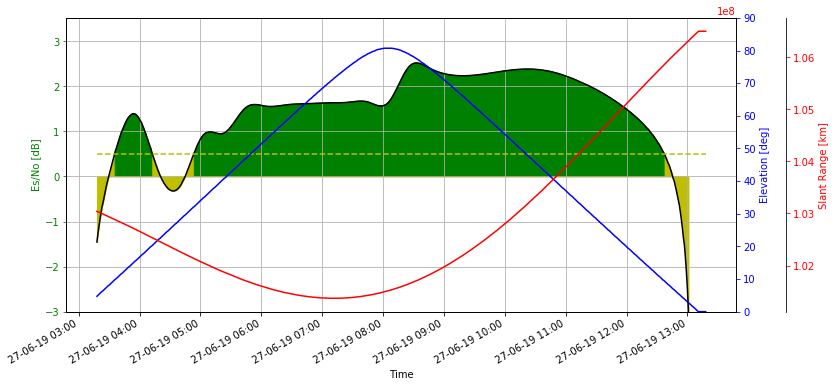}
    \caption{An example output of the linkbudget calculator. The black line shows the overall signal-to-noise ratio. The colour of the fill under the curve corresponds to different download rates, yellow for low bit rate, green for high bit rate. In blue is the elevation of the spacecraft above the horizon from the perspective of the ground-station. In this example the craft passes nearly overhead, at almost 80 degrees above the horizon. In red is the slant range, the direct distance to the space craft. The distance decreases at first as the spacecraft approaches. The complex curve of the overall link-budget is produced through the interaction of these factors and many more.}
    \label{fig:linkbud}
\end{figure}

The link budget calculation is found based on the technical capabilities of the spacecraft antenna \& power in different modes, the inclination, range and orientation of the spacecraft relative to the ground-station, technical characteristics of the ground-station receiver, aspects of the ground-stations local environment and the altitude. The result is something similar to figure \ref{fig:linkbud}.

\subsection{Scheduling environment}
\begin{figure}
    \centering
    \includegraphics[width=\linewidth]{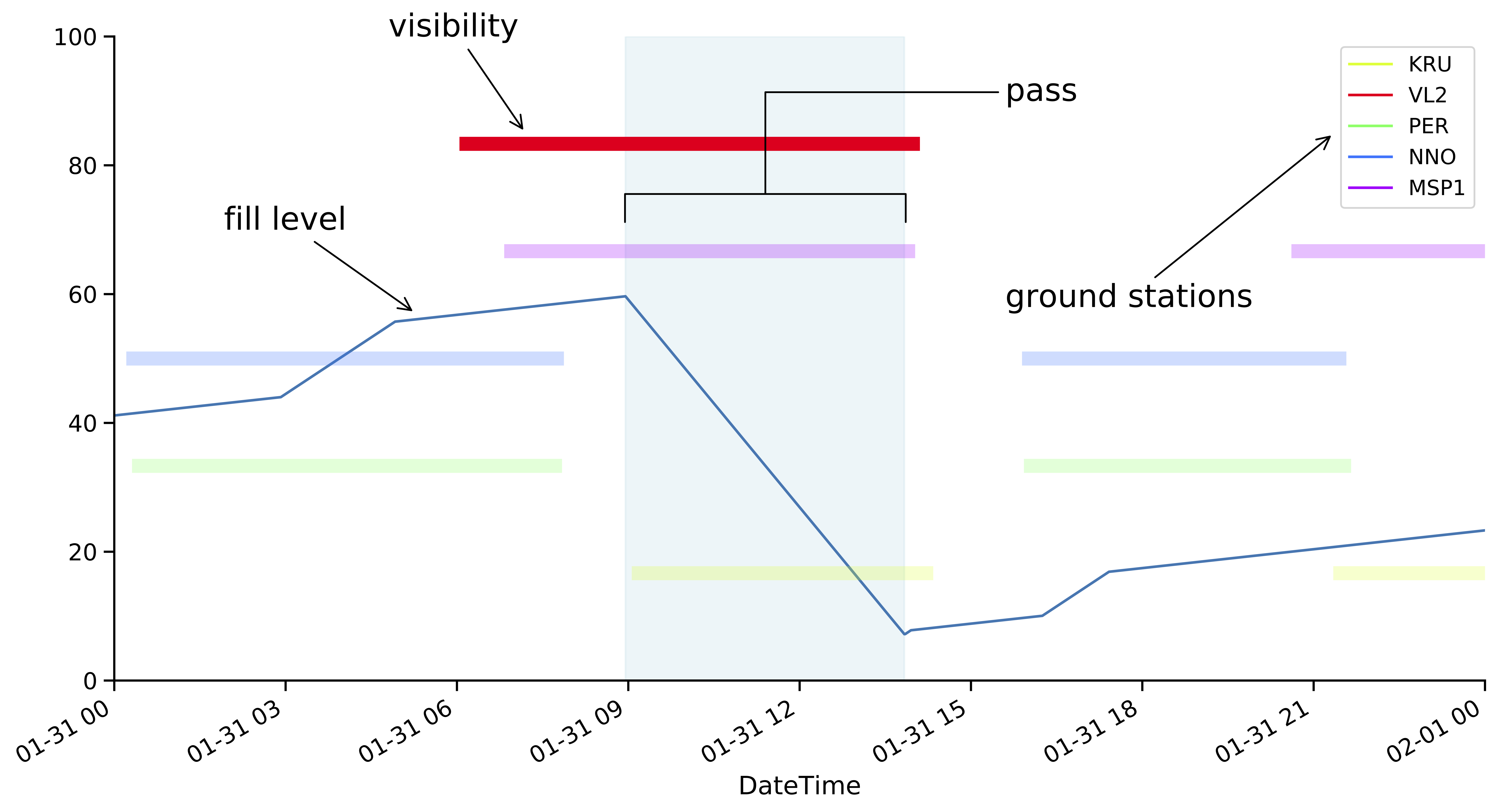}
    \caption{An annotated example of the state of the scheduler for a single spacecraft.}
    \label{fig:scheduler}
\end{figure}

The scheduling environment combines information about when visibilities and different science modes (data production speeds) will occur with link-budgets and booked passes, and propagates the fill-level of each spacecraft over the time period. Altogether this is the 'state' of the schedule, to which we can add (or remove) passes and which can be scored. An annotated example is provided in figure \ref{fig:scheduler}.

\section{Models}
\label{appendix:models}
\subsection{Operator decision-making}
\begin{figure}
    \centering
    \includegraphics[width=0.85\linewidth]{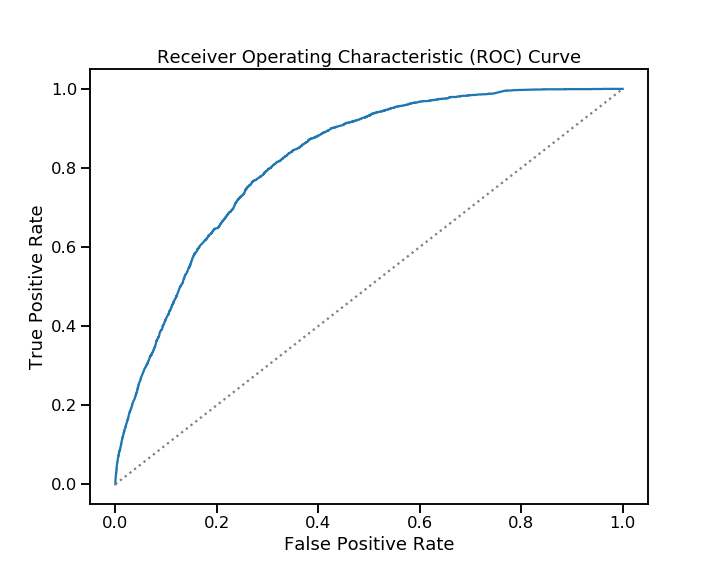}
    \caption{Receiver-operating-characteristic curve for the ensemble. A random classifier would produce something near to the identity line, a perfect classifier would go from (0,0) to (0,1) to (1,1). This model performs very well. The area-under-the-curve is 0.79 which is a strong score, particularly given the reduce context provided. Note that the flat region in the top right from (0.8,1) to (1,1) indicates that the model achieves very high accuracy for those passes it is most certain that were booked. This probably corresponds to passes of exceptional quality that occur when the storage was nearly full.}
    \label{fig:roc}
\end{figure}

A gradient boosting ensemble of 50 decision trees with max depth 3 was trained to predict whether an operator had booked a pass in a visibility given statistical descriptions of the link-budget of the visibility and the fill level of the spacecraft. The descriptions of the link-budget related to the duration (total duration, low bit rate duration, high bit rate duration) and the `quality' of the visibility ($25^{th},50^{th},75^{th}$ and $100^{th}$ percentiles of the signal strength). Note that this representation is agnostic to much of the context used by operators to make decisions, most notably the relative quality of upcoming visibilities and how booking the pass will affect the options for the other craft. Figure \ref{fig:roc} shows the ROC curve for the model. The model achieved an, unexpectedly high, AUROC of 0.79.

\subsection{Schedule beam search}
Beam search is a kind of heuristic tree search that holds a limited number of candidate solutions, the 'beam width'. There are two key stages, branching and pruning. First, each node is expanded to all candidate nodes (possible next moves). A branching heuristic is then used to pre-select a limited number of candidates per node. A second heuristic is then used to evaluate all candidates generated across the nodes and prune to leave just the beam width. This allows for tightly controlled computational cost at each step in the rollout. Our initial solution only used a pruning heuristic that was derived from a combination of scores. In a more developed model we would incorporate the human-modelling, either in combination with quick-to-evaluate scores in a branching heuristic (pre-selection) or, with a more advanced model with access to a realistic context, as part of the pruning heuristic.

\end{document}